\documentclass{article}
\usepackage[utf8]{inputenc}		% For UTF-8 support
\usepackage{xcolor,graphicx}				% Allow for color (annotations)

\def\mmunit{\mathsf{mm}}
\def\nmunit{\mathsf{nm}}
\newcommand{\dd}{{\mathrm d}}
\newcommand{\ex}[1]{{\mathrm e}^{#1}}                 % exponential
\def\rmi{\mathrm{i}}

\begin{document}
\null\vfil\begin{center}{\LARGE Time to Stop Telling Biophysics Students That Light Is Primarily a Wave}\end{center}\medskip\begin{center}{\LARGE
Philip C. Nelson\\ Department of Physics and Astronomy\\ University of Pennsylvania\\
Philadelphia PA USA}\end{center}
\bigskip\noindent \textit{Running head:} Stop telling students light is a wave
\vfil\newpage

\begin{abstract}
Standard pedagogy introduces optics as though it were a consequence of Maxwell's equations, and only grudgingly admits, usually in a rushed aside, that light has a particulate character that can somehow be reconciled with the wave picture. Recent revolutionary advances in optical imaging, however, make this approach more and more unhelpful: How are we to describe two-photon imaging, FRET, localization microscopy, and a host of related techniques to students who think of light primarily as a wave? I was surprised to find that everything I wanted my biophysics students to know about light, including image formation, x-ray diffraction, and even Bessel beams, could be expressed as well (or better) from the quantum viewpoint pioneered by Richard Feynman. Even my undergraduate students grasp this viewpoint as well as (or better than) the traditional one, and by mid-semester they are already well positioned to integrate the latest advances into their understanding. Moreover, I have found that this approach clarifies my own understanding of new techniques.\end{abstract}

The study of light occupies a unique place in the history and current practice of biology, physics, and many other sciences. It is no exaggeration to say that everyone is interested in light and vision, as you can confirm by admitting to a stranger on an airplane that this is a topic that you study. Light and vision are metaphors for insight and wisdom. Visible light imaging is enjoying an extraordinary renaissance, with new techniques arriving nearly every month. Nearly all life is powered by light from the Sun.
So perhaps it is appropriate to spend a moment rethinking what we teach our students---particularly biophysics students---about light.

\section{The ``page 969'' approach}
We typically assign our first-year undergraduates a Physics textbook that weighs several kilograms. I just looked at a typical specimen. After more than 900 pages, this book begins to discuss ``light.'' There is a graphic depicting the electric and magnetic fields in a wave, but very quickly that approach is discarded: After just a couple of pages we find the magisterial transition, ``Huygens' principle can be justified from a detailed study of the behavior of waves in Maxwell's equations, although we shall not do so here.'' There follows a mixture of wavelets and ray diagrams. Then a later chapter on ``Interference'' reintroduces waves. Yet another chapter on ``Diffraction'' arrives, after 150 pages, at the Rayleigh criterion for the diffraction limit on resolution. Then the discussion of optics is over without having mentioned the particulate character of light. \emph{That} aspect arrives later still, in a chapter focused on blackbody radiation---a phenomenon whose relevance is not obvious to a life science student.

Thus, even if a first-year course makes it to page 1000 of this book, our students leave with a mashup of different viewpoints (ray optics, wavelets, physical optics), and a vague sense that it all comes from the fact that light ``is'' a wave (except when it isn't).
These students eventually arrive in our labs, where they confront the arsenal of modern biophysics technique, including fluorescence microscopy, two-photon imaging, resonance energy transfer, localization microscopy---none of which will make sense to an undergraduate who thinks that light is a wave. It's no wonder that the general attitude of life-science students to Physics is that it's a meaningless filter, a hoop they must jump through and then forget.

\section{The ``day 1'' approach}
Eventually I asked myself: Why pretend that light is primarily just a wave? What does that stance explain that life-science students care about? Or bluntly, \emph{what do you gain} by starting with the 18th--19th century viewpoint?
The answer appears to be, ``Very little.'' Sure, Maxwell's equations are a fabulously accurate approximation if we are designing a radio telescope. But much of what today's biophysics students urgently need to understand involves \emph{one photon} at a time, interacting with \emph{one molecule.} 

So I decided instead to dive in on day one, putting the particulate aspect of light front and center, and to keep it that way while still obtaining the classic results from optics. I start by showing my class the time series of signals that came from an avalanche photodiode, or other sensitive detector, at different levels of steady, low illumination. We observe that (a) the signals come as discrete blips; (b) the blips are always the same regardless of the intensity of the illumination; (c) the blips arrive at random; (d) the mean arrival rate encodes the intensity. Next I demonstrate the photoelectric effect, establishing that (e) each blip is the deposition of a lump of \emph{energy,} whose magnitude depends on where the light sits in the spectrum (``color''). Finally, data from a detector array shows that (f) each blip is highly localized in space; no two pixels of the array ever fire simultaneously, quite unlike what we would expect if a wave passed by and jiggled every electron in its path in unison.

The six points just mentioned already enable us to understand qualitatively many cutting-edge biophysical phenomena. Casting everything in terms of energy lets us draw upon students' understanding of chemistry (molecular energy levels) to make sense of fluorescence and its Stokes shift, two-photon excitation, and even resonance energy transfer. The one photon/one molecule aspect lets us say interesting things about photosynthesis and even optogenetics. We can also see why we only get skin cancer (genetic damage) from UV light, not indoor lighting, why premature infants are treated with blue light to break down excess bilirubin, and so on. Students are now engaged; they have heard these topics out on the street. They are now prepared to go the distance as things get a bit more abstract.

How can it be intellectually honest to build on the assumed black box of quantized energy levels, without explaining it? Perhaps only a physicist would even worry about that. After all, everything sits on \emph{some} bedrock. My point here is that if we make just a \emph{few} assumptions, and they give a \emph{broad range} of relevant, quantitative, and confirmed predictions, then we are doing science, even if we will later revisit those assumptions and obtain them from still more general foundations. Moreover, if we stay close to observable phenomena, especially those that can be demonstrated live in a classroom or lab, then assumptions, while provisional, are allowed. This is after all how we make progress in our research. Nor should we let a century or so of pedagogical tradition blind us to the gaps in the usual approach, for example the unproven step where we wave a wand and introduce Huygens' principle, or more fundamentally the very idea of ``electromagnetic field.''

\section{Yes, but\ldots}
Of course, eventually we must address the many \emph{apparently} wavelike phenomena of light. Actually, however, there is nothing wavelike about even two-slit diffraction; I show classic videos demonstrating how it (and also electron diffraction patterns) build up one localized blip at a time. What we need in order to understand such patterns, then, is a rule that lets us calculate the \emph{probability density function} for where the next localized packet of energy (photon) will arrive. So I walk students through Richard Feynman's prescription for finding this function by summing a probability amplitude over allowed paths \cite{Bfeyn85a}.

Is this approach abstract and conceptually challenging? Yes, certainly, but so is organic chemistry. Is it \emph{unnecessarily} abstract and challenging? I would say no: After a century, this approach (or one equivalent to it) is still the \emph{only known way} to reconcile the particle-like and wave-like aspects of light. It is a physical hypothesis; we cannot derive it from any deeper and more palatable layer of reality. And there is a payoff: For example, the new technique of interferometric photoactivation localization microscopy  \cite{Case:2015p14837,%VanEngelenburg:2014p12085,
Shtengel:2014p13997,%Kanchanawong:2010p6346,
Shtengel:2009p6347} simultaneously makes use of the particulate character of light (for superresolution imaging in the $xy$ plane) and its wave character (for the $z$ coordinate of each fluorescent molecule).

Fortunately, although conceptually challenging, this viewpoint is \emph{mathematically} straightforward to implement in many situations of direct biophysical interest. After working out diffraction from a pair of of thin slits, my students can readily calculate the pattern to be expected from slits of  variable width, and see mathematically the transition from a regime where light seems to travel on straight lines (wide slit) to one where the ray approximation fails badly (narrow slit). Then I bring in a laser setup and we see the same thing directly with our eyes.
We can now add one more element (slowing of light by a medium), and find that the stationary-phase principle predicts refraction. That brings topics such as total internal reflection into reach. 

\section{Imaging}
We are now in a position to discuss image formation. Following Feynman's  discussion \cite{Bfeyn85a}, we ask ``What kind of optical element can adjust the phases of light paths so that nearly all the light from a point source arrives at a single point on a detector array?'' The answer of course turns out to be a lens, but now we can understand not only traditional (lens-shaped) lenses, but also gradient-index lenses (which can even be \emph{flat}). The famous formulas for thin lenses are no more difficult to obtain in this approach than in the traditional way.

As a bonus, we also get a unified approach to the Rayleigh diffraction ``limit,'' instead of having to throw it in as a disconnected later chapter: For finite aperture size, the stationary-phase approximation is of limited validity, which we can assess quantitatively. When we do the calculation without that approximation, diffractive blur appears.

Because we have not told any fibs about light being a wave, we now also have a self-consistent picture applicable to today's world of single-molecule imaging, where the particulate character of light is essential, and even to fluorescence imaging, whose utility hinges on the Stokes shift phenomenon. Photoswitching is now at least plausible, thanks to the connection made earlier to discrete molecular states. 

The probabilistic viewpoint also motivates a search for an inference technique that can find the center of a point-spread function to greater accuracy than its width. Combining those last two ideas brings students directly into superresolution localization microscopy---all within a single, self-consistent framework.

\section{Vistas}
\subsection{Bessel beams}
Here is one example topic among many that could be cited, taken from current headlines. Light-sheet microscopy offers many benefits, but its use in thick samples is limited by the diffractive spreading of an ordinary beam of light: The illuminated region is too wide. Recently, however, this limitation has been addressed by an application of ``Bessel beam'' illumination \cite{Legant:2016p14728,Gao:2014gi}. A Bessel beam is generally explained as a remarkable solution to the Maxwell equations for which the central maximum does not spread. Crucially, an approximate version of such a beam is readily realizable in the lab, and it, too, spreads much less than an ordinary Gaussian beam of the same initial diameter \cite{Durnin:1987p15310}. 

\begin{figure}
\begin{center}	\includegraphics{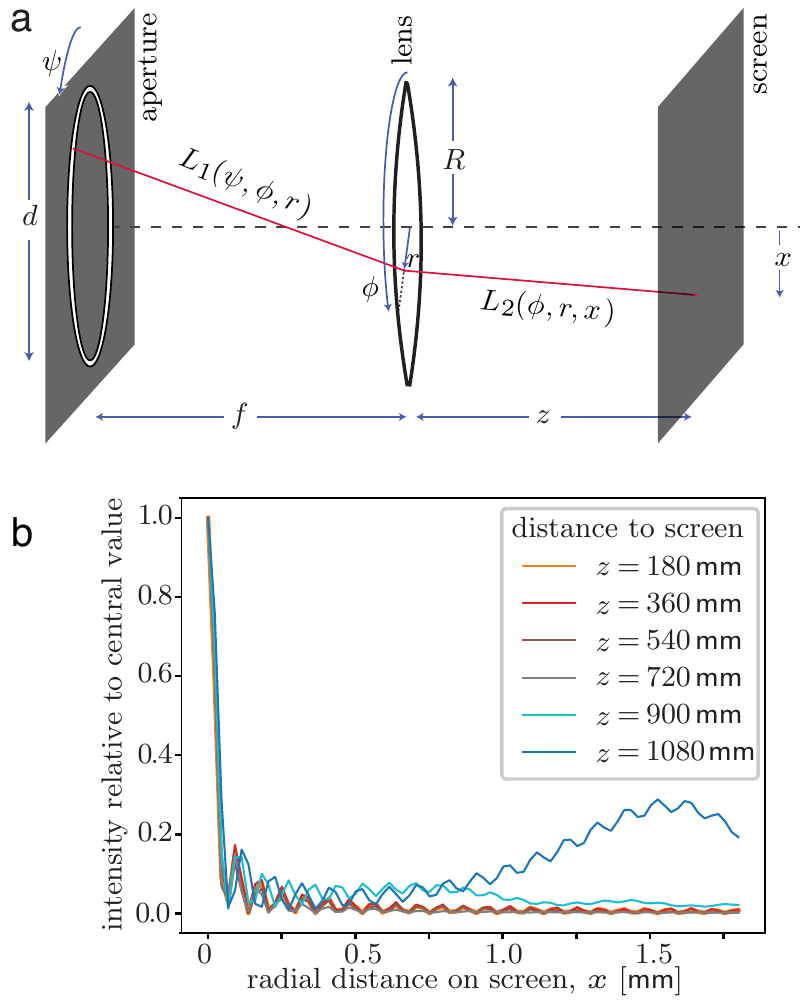}\end{center}
	\caption{(\textsf a) {\small Schematic of a setup to generate a Bessel beam. A lens is placed one focal length $f$ away from a narrow, annular aperture, which is illuminated by a laser. The pattern of illumination is observed on a projection screen a further distance $z$ away. Each of the photon paths considered consists of straight-line segments, and is characterized by position on the aperture ($\psi$), and the position at which it passes through the lens ($r$, $\phi$). Not to scale.}
(\textsf b)~{\small Numerical integration over these three variables leads to a predicted light intensity pattern that remains tightly confined to a central beam, in this case out to a distance of nearly a meter from the lens. The calculation assumed ring aperture diameter $d=5\,\mmunit$, $f=305\,\mmunit$, wavelength $633\,\nmunit$, and lens diameter $2R=7\,\mmunit$, as in the original demonstration \protect\cite{Durnin:1987p15310}. Each curve was separately normalized to peak at 1.
}
\label{f:bbe}}
\end{figure}
The experimental setup is simplicity itself: A lens is placed one focal length away from an annular slit aperture.
Unfortunately most of us (certainly me) would find it daunting to explain the ensuing behavior of a Bessel beam by solving Maxwell's equations in the appropriate paraxial regime \cite{Bmill10b,Bsimo16a}. However, the corresponding path integral calculation is straightforward. We consider photon paths that consist of two  segments: aperture$\to$lens, and lens$\to$detector. We then integrate a phase factor over all such paths, that is, over the three variables $\phi,\ \psi,\ r$ shown in Fig.\ \ref{f:bbe}a:
$$f(x,z) = \int_0^{2\pi}\dd\psi\int_0^{2\pi}\dd\phi\int_0^R\dd r\,ABC.$$%
In this formula, $A(\psi,\phi,r)=(1/L_1)\ex{2\pi\rmi L_1/\lambda}$ is the contribution to the probability amplitude from the first part of the photon path. Similarly, $C(\phi,r,x,z)=(1/L_2)\ex{2\pi\rmi L_2/\lambda}$ comes from the last part. The lens contributes $\ex{(2\pi\rmi/\lambda)(r^2/(2f))}$ because glass slows light down, and the distance traversed in the lens depends on $r$. The variables $\psi,\phi,r,R,x$, and $z$ are defined in the figure and its caption. Numerically
computing the modulus squared of the complex function $f$ gives the intensity profile (Fig.\ \ref{f:bbe}b), which indeed spreads much more slowly than light emerging from a pinhole. (A similar but easier calculation gives the usual Airy function for the pinhole case, again without solving any differential equation.)
Durnin and coauthors gave a formula for the critical distance at which the beam begins to spread, which in this case gives $2fR/d\approx900\,\mmunit$ in agreement with their experiment and the present calculation \cite{Durnin:1987p15310}.

\subsection{Other biophysical phenomena}
Summing over paths is also an elegant approach to working out interference effects such as iridescent colors in insect wings, as well as thin-film reflection in optical instruments. Maintaining a photon viewpoint lets us transition to other physiological matters, such as photoreception, and from there to color vision and the single-photon sensitivity of vertebrate vision, all in a single semester.

\subsection{Unity}
Scientists like unity, a small number of principles that explain diverse phenomena. From that viewpoint, it is breathtaking to find that, although the quantum rules for light are crazy, nevertheless the \emph{same crazy rules} apply to electrons---matter---as well. Electrons can diffract; in regimes where diffraction is negligible, they follow paths dictated by a stationary-phase principle, just like light. Moreover, Feynman's rule, adapted for electrons, predicts the key fact of quantized energy levels without additional hypotheses! Even students who have taken a full year of physical chemistry, or quantum mechanics, have not been told this. Making it the starting point of our study of electrons (the ``page 1'' approach) lets them connect the conceptual islands in a fruitful way.

\section{When and why to introduce the Maxwell equations}
Realistically, your Physics department will continue teaching electromagnetism in the same way until the spherical cows come home. This is not a bad thing; biophysicists need to know about intermolecular forces and so on, and the subject also serves as basic training for some useful mathematical techniques. I am suggesting that initially, we discuss optical phenomena \emph{independently} of electromagnetism, because (a)~the quantum aspects are so central to biophysics, and do not fit well into a 1-semester course on classical electromagnetism;
and (b)~anyway, in many Physics departments optics has already shriveled into a small footnote to the first-year course (or disappeared altogether).
Because the photon story does not rely on any partial differential equations, you can even tell it \emph{before} classical electromagnetism should you choose.

Certainly for more technical problems it is also valuable to know that the photon probability amplitude obeys a partial differential equation, which is often the best approach to take for Gaussian beams propagating through optical elements \cite{Bsieg86a}. And
of course, light \emph{does} have {something} to do with electricity and magnetism---and that surprise is a central part of our intellectual heritage. Although we can go a long way in biophysics without explicitly mentioning this connection, eventually an advanced student will want to know about it, for example to understand polarization effects. Even these effects, however, often manifest themselves at the single-molecule level in biophysics, so we still cannot treat them adequately by using the wave picture. For example, many invertebrates possess polarization vision. Their visual pigment isomerizes upon absorption of a single photon, and a single photoisomerization suffices to elicit a neural response, so again the biophysics is intrinsically quantum-mechanical. In the realm of instrumentation, polarization total internal reflection microscopy, too, is often used to observe the absorption and emission of single photons by single molecules. 

To understand light at this level, we must quantize the electromagnetic field. The role of Maxwell's equations is then seen as motivating the correct starting point for this construction---a program that should probably be deferred until graduate school. Luckily, vertebrate vision is almost totally insensitive to polarization, and elementary optical instruments are also traditionally studied neglecting this aspect.

\section{Yes, you can}
Any reader who has come this far may well be saying, ``But you could never take that approach with the real undergraduatesthat we get.'' When I hit that point, eventually I asked myself, ``How much working understanding are our students getting from the \emph{traditional} approach?'' They can perhaps do some potted problems, but are they getting the basis on which to understand new methods? From that perspective, the approach advocated in this essay may make more sense.

Moreover, consider a trend at my own and many other universities. Bioengineering is growing rapidly---those students are keenly interested in these topics, and they are not afraid of math, if it \emph{builds a coherent picture} and remains \emph{rooted in practical issues,} for example, seeing what was previously invisible.

Another likely response is ``I don't teach introductory courses, nor do I have any leverage on those who do.'' Do not worry. Your students have all \emph{forgotten} their first-year classes anyway; it's not too late to tell them things they need. I found I could present the ideas in this article in a one-semester class for third-year undergraduates in any quantitative science major \cite{Bnels17a}, but even grad students will find much here that is new to them.

Finally, for my own selfish reasons I wanted a framework in which I could understand the things I hear at the Biophysical Society meetings and read in the \textsl{Biophysical Journal,} and connect them to existing concepts already in my head. I believe that in the long run, this imperative is itself one of the values we are trying to transmit to our students.

\section*{Acknowledgments}
This work was partially supported by the United States 
National Science Foundation under Grant PHY--1601894. Some of the work was done at
the Aspen Center for Physics, which is supported by NSF grant PHY--1607611.

\bibliographystyle{unsrt}
% \bibliography{AAPTbibdesk,AAPTpapers}

\begin{thebibliography}{10}

\bibitem{Bfeyn85a}
R~P Feynman.
\newblock {\em {QED}: {T}he strange theory of light and matter}.
\newblock Princeton Univ. Press, Princeton NJ, 1985.

\bibitem{Case:2015p14837}
L~B Case, M~A Baird, G~Shtengel, S~L Campbell, H~F Hess, M~W Davidson, and C~M
  Waterman.
\newblock {Molecular mechanism of vinculin activation and nanoscale spatial
  organization in focal adhesions}.
\newblock {\em Nat. Cell Biol.}, 17(7):880--892, 2015.

\bibitem{Shtengel:2014p13997}
G~Shtengel, Y~Wang, Z~Zhang, W~I Goh, H~F Hess, and P~Kanchanawong.
\newblock {Imaging cellular ultrastructure by PALM, iPALM, and correlative
  iPALM-EM}.
\newblock {\em Meth. Cell Biol.}, 123:273--294, 2014.

\bibitem{Shtengel:2009p6347}
G~Shtengel, J~A Galbraith, C~G Galbraith, J~Lippincott-Schwartz, J~M Gillette,
  S~Manley, R~Sougrat, C~M Waterman-Storer, P~Kanchanawong, M~W Davidson, R~D
  Fetter, and H~F Hess.
\newblock {Interferometric fluorescent super-resolution microscopy resolves
  {3D} cellular ultrastructure}.
\newblock {\em Proc. Natl. Acad. Sci. USA}, 106(9):3125--3130, 2009.

\bibitem{Legant:2016p14728}
W~R Legant, L~Shao, J~B Grimm, T~A Brown, D~E Milkie, B~B Avants, L~D Lavis,
  and E~Betzig.
\newblock {High-density three-dimensional localization microscopy across large
  volumes}.
\newblock {\em Nat. Methods}, 13(4):359--365, 2016.

\bibitem{Gao:2014gi}
L~Gao, L~Shao, B-C Chen, and E~Betzig.
\newblock {3D live fluorescence imaging of cellular dynamics using Bessel beam
  plane illumination microscopy.}
\newblock {\em Nat. Protoc.}, 9(5):1083--1101, 2014.

\bibitem{Durnin:1987p15310}
J~Durnin, J~Miceli, and J~Eberly.
\newblock {Diffraction-free beams}.
\newblock {\em Phys. Rev. Lett.}, 58(15):1499--1501, 1987.

\bibitem{Bmill10b}
P~W Milonni and J~H Eberly.
\newblock {\em Laser physics}.
\newblock Wiley, 2010.

\bibitem{Bsimo16a}
D~S Simon.
\newblock {\em A guided tour of light beams: {F}rom lasers to optical knots}.
\newblock Morgan and Claypool, San Rafael CA, 2016.

\bibitem{Bsieg86a}
A~E Siegman.
\newblock {\em Lasers}.
\newblock University Science Books, Mill Valley, CA, 1986.

\bibitem{Bnels17a}
P~Nelson.
\newblock {\em From photon to neuron: {L}ight, imaging, vision}.
\newblock Princeton Univ. Press, Princeton NJ, 2017.

\end{thebibliography}

\end{document}